\documentclass[
    final            
  ,numberedheadings 
  ]
  {aipproc}

\layoutstyle{6x9}

\begin{document}

\title{Quantum metrology with rotating matter waves in different geometries}


\author{J. A. Dunningham}{
  address={School of Physics and Astronomy, University of Leeds, Leeds LS2 9JT, United Kingdom}
}

\author{J. J. Cooper}{
  address={School of Physics and Astronomy, University of Leeds, Leeds LS2 9JT, United Kingdom}
}

\author{D. W. Hallwood}{
  address={Institute of Natural Sciences, Massey University, Private Bag 102904, Auckland, New Zealand}
}

\begin{abstract}
A promising practical application of entanglement is metrology, where quantum states can be used to make measurements beyond the shot noise limit. 
Here we consider how metrology schemes could be realised using atomic Bose-Einstein condensates (BECs) trapped in different potentials. In particular, we show that if a trapped BEC is rotated at just the right frequency, it  can undergo a quantum phase transition characterised by large-scale entanglement spreading across the system. This simple process of stirring can generate interesting quantum states such as macroscopic superpositions of all the atoms flowing in opposite directions around a ring-shaped potential.  We consider different trapping potentials and show how this leads to different entangled states. In particular, we find that by reducing the dimensionality of the system to one or two dimensions, it is possible to generate entangled states that are remarkably robust to the loss of atoms and so are ideally suited to precision measurement schemes.
\end{abstract}

\maketitle


\section{Introduction}

One of the most exciting recent applications of quantum information theory has been to the field of metrology. A number of theoretical proposals \cite{Giovannetti2004,Holland1993a,Dowling1998a,Campos2003a} have demonstrated that it is possible to use entanglement to surpass the so-called shot-noise limit. This limit is due to unavoidable statistical fluctuations that arise when discrete independent events are used to make a measurement. It is characterized by a precision that scales as $1/\sqrt N$ where $N$ is the number of discrete events recorded, such as tosses of a coin or particles detected.

If, instead, we remove the constraint that the subsystems are independent of one another and allow them to be entangled,  then it is possible to reach the Heisenberg limit whereby the measurement precision scales as $1/N$. There have even been some proposals for schemes that exploit nonlinearities to go beyond the Heisenberg limit \cite{Boixo2008a, Boixo2009a}. However, it is now thought that these schemes misinterpret the resources required and so the Heisenberg limit is indeed fundamental \cite{Kok}.

Metrology schemes are often based on interferometers because they offer unrivalled precision in the measurement of phase shifts or path length differences. As well as their familiar linear forms, interferometers can also be successfully implemented in ring geometries to make accurate measurements of angular momentum. These interferometric gyroscopes form a key component of many modern navigation and stabilisation systems.
They work by exploiting the different path lengths experienced by light as it propagates in opposite directions around a rotating ring. 
For instance, in the Sagnac geometry \cite{Sagnac1913} photons are put into a superposition of travelling in opposite directions around a ring and, when rotated, the two directions acquire different phases due to their different path lengths. This phase difference is directly related to the rate of rotation and can be measured by recombining the two components at a beam splitter. 

The precision of gyroscopic devices can be improved further still with the use of entangled atoms rather than photons. The precision with which a Sagnac interferometer can measure a phase difference is  \cite{Dowling1998}
\begin{eqnarray}
\Delta\phi = \frac{4\pi}{\lambda v}\Omega . A,
\end{eqnarray}
where $\lambda$ is the particle wavelength, $v$ is the particle speed, $\Omega$ is the angular velocity of the ring, and $A$ is the enclosed area. For equal values of $\Omega$ and $A$, the ratio of the precision for atoms with mass $M$ to photons with frequency $\omega$ is $Mc^2/\hbar\omega$, which for reasonable parameters is about $10^{11}$. So, even though the particle fluxes are likely to be much lower for atoms and the enclosed areas smaller, the very large enhancement factor for atoms means that they are likely to offer considerable advantages over photons.

The use of entangled atomic states to make precision measurements of rotations has received a lot of recent attention. It is possible to achieve sub-shot-noise measurements by using number-squeezed atomic states as the input to an interferometer and so there has been a lot of research into the generation and uses of these states \cite{Orzel2001, Li2007, Esteve2008,  Haine2009}.  Several proposals have already been made to use different non-classical  atomic states to make general phase measurements with sub-shot noise sensitivities \cite{Bouyer1997, Dunningham01a, Dunningham04, Pezze2005, Pezze2006, Pezze2007, Pezze2009}.  
It has also been shown that uncorrelated atoms can also achieve sub shot-noise sensitivities of rotational phase shifts \cite{Search2009} by using a chain of matter wave interferometers, or a chain of gyroscopes.  

It is clear that there is a link between entanglement or squeezing and the measurement precision that can be achieved, however the precise relationship is not obvious. While it can be shown  that a NOON state of the form,
\begin{eqnarray}
|\psi\rangle = \frac{1}{\sqrt2}\left(|N\rangle_a|0\rangle_b + |0\rangle_a|N\rangle_b \right),
\end{eqnarray}
where the two kets represent Fock states of two states $a$ and $b$ respectively,
gives the best possible precision (i.e. it saturates the Heisenberg limit) in the lossless case, the same is not true when we account for the loss that will inevitably be present in any realistic system. The problem is that NOON states are known to be extremely fragile. The loss of a single particle destroys the superposition. So while, they have excellent short-term sensitivity, their long-term stability is very poor. This means that they can only be used to make very short timescale comparisons, which degrades their overall precision. Think of frequency standards: the longer we can compare our system with the standard, the more accurately we can detect any difference in their frequencies. So there is an important trade-off between the entanglement in a system and the robustness of the state to decoherence.

We can quantify how accurately a state can measure an unknown parameter with the quantum Fisher information and its associated Cram\'er-Rao bound. Quantum Fisher information is defined as,
\begin{eqnarray}
F_Q = {\rm Tr}[\rho(\phi)A^2],
\end{eqnarray}
where $\rho(\phi)$ is the density matrix of the system, $\phi$ is the parameter to be measured, and $A$ is the symmetric logarithmic derivative defined by,
\begin{eqnarray}
\frac{\partial\rho(\phi)}{\partial\phi} = \frac{1}{2}[A\rho(\phi) + \rho(\phi)A].
\end{eqnarray}
In the eigenbasis of $\rho(\phi)$ this is,
\begin{eqnarray}
(A)_{ij} = \frac{2}{\lambda_i + \lambda_j}[\rho'(\phi)]_{ij},
\end{eqnarray}
where $\lambda_{i,j}$ are the eigenvalues of $\rho(\phi)$ and $\rho'(\phi) = \partial\rho(\phi)/\partial\phi$. If $\lambda_{i} + \lambda_j =0$ then $(A)_{ij} =0$. 
The Fisher information is related to the uncertainty of the measurement via the Cram\'er-Rao bound, $\Delta\phi \geq 1/\sqrt{F_Q}$. For a pure state, $|\psi(\phi)\rangle$, we have,
\begin{eqnarray}
F_Q = 4\left[ \langle \psi'(\phi)|\psi'(\phi)\rangle - |\langle \psi'(\phi)|\psi(\phi)\rangle |^2\right], \label{fisherpure}
\end{eqnarray}
where $|\psi'(\phi)\rangle = \partial |\psi(\phi)\rangle/\partial\phi$.
Fisher information will serve as our yardstick for comparing the performances of different entangled states in metrology.

In this paper we will discuss how a variety of entangled states that are useful for metrology can be created simply by stirring Bose-Einstein condensates (BECs) trapped in potentials with different geometries. We start with the case of NOON states and then show how, by modifying the potential and dimensionality of the system, it is possible to create different entangled states that perform better in the presence of decoherence. The general study of the relationship between entanglement, phase transitions, and the geometry of the trapping potential is an exciting new area of research that is just starting to open up. This development has been motivated, in part, by recent outstanding experimental advances in the ability to trap atoms in a wide variety of time-varying potentials.

\section{Lattice and continuous ring potentials}
\label{sec:lattice_ring}
\begin{figure}
  \includegraphics[height=.3\textheight]{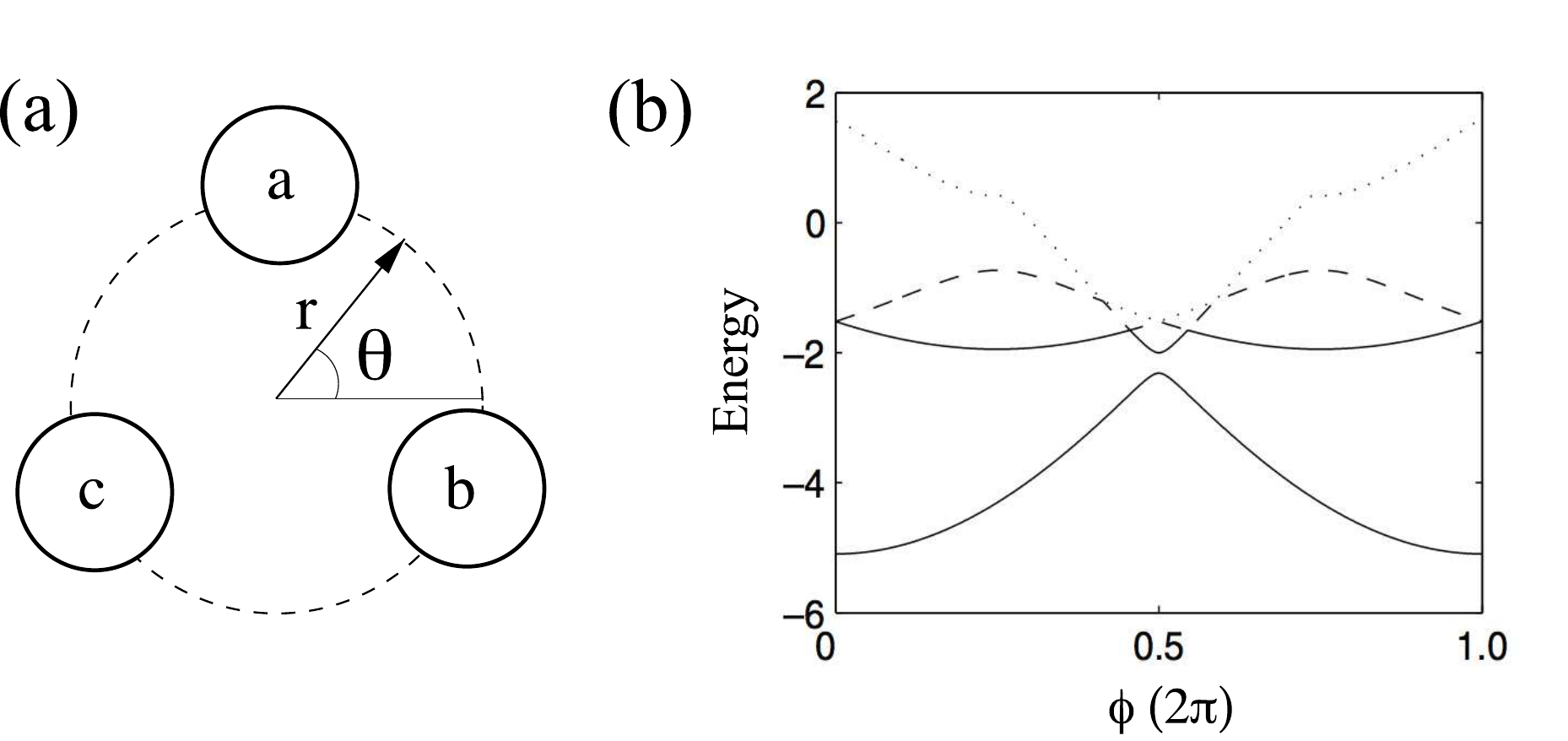}
  \caption{(a) The three site ring lattice. Atoms are trapped at sites $a$, $b$, and $c$ and can tunnel between adjoining sites. (b) Energy levels (in units of J) plotted as a function of the rotation, $\phi$, for $J=U=1$ and $N=3$. A clear anti-crossing between the two lowest energy levels is seen for $\phi=\pi$.} \label{fig:3site}
\end{figure}

The first trapping geometry that we will consider is a lattice of potential nodes in a ring configuration. For simplicity, we will consider a lattice consisting of just three sites as shown in Figure~\ref{fig:3site}(a) since this is the minimum number required for a ring. The different sites are coupled to one another via quantum mechanical tunnelling through the barriers separating them. The atoms on any given site also experience a nonlinearity due to collisions with other atoms on that site. This trap geometry has already been experimentally demonstrated \cite{Boyer2006, Henderson2009}.

Assuming that the system is sufficiently cold that we can consider only one trap level at each site, the system can be described by the Bose-Hubbard Hamiltonian. The rotation of the sites can be incorporated into the Bose-Hubbard Hamiltonian \cite{Jaksch1998} simply by including appropriate phase factors in the coupling terms. This gives the `twisted' Hamiltonian, 
\begin{eqnarray}
H &=& -J\left[ e^{i\phi/3}\left( a^\dagger b + b^\dagger c + c^\dagger a\right) + h.c.\right]  + U \left( {a^\dagger}^2 a^2 + {b^\dagger}^2 b^2 + {c^\dagger}^2 c^2\right),
\end{eqnarray}
where $a$, $b$ and $c$ are the annihilation operators of atoms in the three sites,
$U$ is the on-site interatomic interaction strength and $J$ is the coupling 
between adjacent lattice sites due to tunnelling through the barrier.
The phase factors in the coupling
terms are known as Peierls phase factors.

It is convenient to rewrite the Hamiltonian in the quasi-momentum basis $\{ \alpha, \beta, \gamma \}$ 
\begin{eqnarray}
&&\alpha = { 1 \over \sqrt{3} } (a + b + c),\nonumber\\ 
&&\beta = { 1 \over \sqrt{3} } (a + b e^{i 2 \pi / 3} + c e^{i 4 \pi / 3}),
\nonumber \\
&&\gamma = { 1 \over \sqrt{3} } (a + b e^{- i 2 \pi / 3} + c e^{- i 4 \pi / 3}).
\label{eq:quasimodes}
\end{eqnarray}
These new basis states respectively correspond to zero flow, one quantum of clockwise flow, and one quantum of anticlockwise flow, where throughout this paper we will use the convention that a positive phase variation corresponds to clockwise flow.

In this basis, the twisted Hamiltonian has the form
\begin{eqnarray}
H & = & - J \{ (2 \alpha^{\dagger} \alpha -
\beta^{\dagger} \beta - \gamma^{\dagger} \gamma) \cos(\phi/3) + \sqrt{3}
(\beta^{\dagger} \beta - \gamma^{\dagger} \gamma) \sin(\phi/3) \} \nonumber \\
&+& \frac{U}{3} \{ \alpha^{\dagger}{}^2 \alpha^2 + \beta^{\dagger}{}^2 \beta^2 + \gamma^{\dagger}{}^2 \gamma^2+ 4(\alpha^{\dagger} \alpha \beta^{\dagger} \beta + \alpha^{\dagger} \alpha \gamma^{\dagger} \gamma  \nonumber \\
&+&   \beta^{\dagger} \beta\gamma^{\dagger} \gamma ) 
+ 2( \alpha^2 \beta^{\dagger} \gamma^{\dagger} + \beta^2
\alpha^{\dagger} \gamma^{\dagger} + \gamma^2 \alpha^{\dagger} \beta^{\dagger} +
h.c.) \}. \label{flowham}
\end{eqnarray}
The full energy spectrum of the system can be found by diagonalising this Hamiltonian. However, we can gain a lot of insight about
the behaviour of the system simply from inspection. When there are no interactions ($U=0$) the second and third line of the Hamiltonian vanish and when there is no rotation ($\phi=0$) we see that the ground state is the non-rotating state, $\alpha$, and the other two states are degenerate, as we would expect. If we were to increase the rotation rate to $\phi = 2\pi$, the ground state would be all the atoms in $\beta$, i.e. all rotating with one quantum of angular momentum. In between these values, i.e. when $\phi=\pi$, $\alpha$ and $\beta$ become degenerate ground states. However, if we include interactions, we see from (\ref{flowham}) that the different modes are now coupled. This serves to lift the degeneracy and causes an anticrossing at  $\phi=\pi$. This can be seen in Figure~\ref{fig:3site}(b), where the energy levels are plotted as a function of $\phi$ for $J=U=1$ and three atoms.

To a very good approximation, the ground state at the critical rotation rate ($\phi=\pi$) is a symmetric superposition of all the atoms in $\alpha$ and all in $\beta$, i.e. a macroscopic superposition or NOON state,
\begin{eqnarray}
|\psi\rangle = \frac{1}{\sqrt2}\left(|N\rangle_\alpha|0\rangle_\beta + |0\rangle_\alpha|N\rangle_\beta \right).
\end{eqnarray}
This is an intriguing result because it means that, simply by stirring a Bose condensate at just the right rate, it is possible to generate a highly entangled state.

Strictly speaking, when the barrier heights are equal, we only get a NOON state by this method when we have commensurate filling, i.e. the total number of atoms is an integer multiple of the number of lattice site. However, it has been shown that we can get around this problem simply by ensuring that the barrier heights are slightly different \cite{Hallwood2006a}. 

\begin{figure}[b]
  \includegraphics[height=.25\textheight]{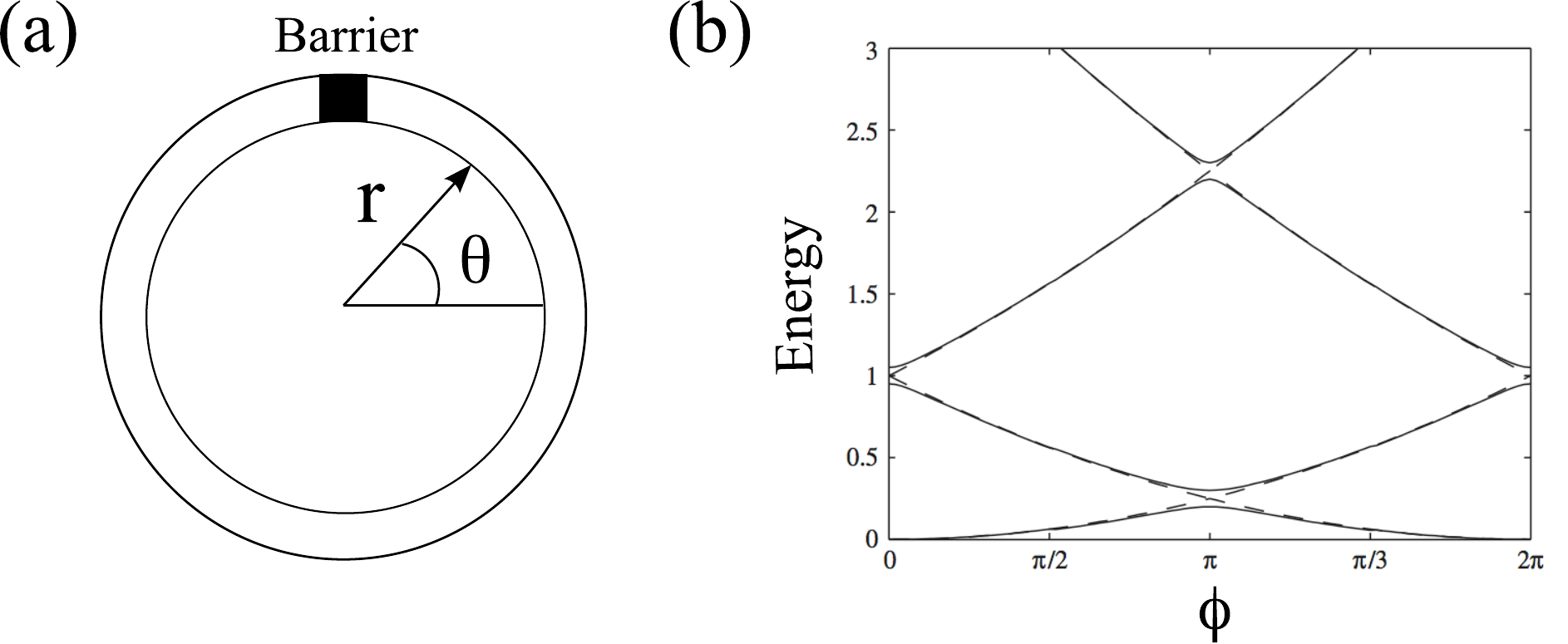}
  \caption{(a) The continuous loop potential with a small barrier that the atoms can tunnel through. (b) Energy levels (in units of $\hbar^2/(2Mr^2)$, where $M$ is the mass of an atom and $r$ is the radius of the loop) plotted as a function of the rotation, $\phi$. The dashed lines are for the case without the barrier and the solid lines are the energy levels when the barrier is included. An anti-crossing between the two lowest energy levels is seen for $\phi=\pi$, i.e. when the loop is rotated at a rate corresponding to half a quantum of angular momentum per particle.} \label{fig:continuous}
\end{figure}

Very similar results have also been demonstrated for atoms trapped in a continuous loop with a small barrier at one point \cite{Hallwood2007a} as shown in Figure~\ref{fig:continuous}(a). The atoms can tunnel through the barrier and the loop is rotated. The energy levels for this system with and without the barrier present are shown in Figure~\ref{fig:continuous}(b). At a rotation rate corresponding to a phase $\phi =\pi$ around the loop (i.e. when the loop is rotated at a rate corresponding to half a quantum of angular momentum per atom) there is an anti-crossing for the case that the barrier is present. As in the case of the lattice, the ground state at this critical frequency is a macroscopic superposition of all the atoms not rotating and all rotating with one unit of angular momentum.

Both the lattice and continuous loop geometries discussed above could be used as entanglement enhanced gyroscopes. The general idea is as follows \cite{Dunningham2006a, Cooper2010a}. We begin with a BEC trapped in a non-rotating version of one of these potentials. We then abiabatically increase the rate of rotation until we reach the critical frequency. At this point, the system should be in the ground state, which is a NOON state of rotating and non-rotating components. Next we make a sudden (non-adiabatic) change to the rotation frequency and allow the system to evolve for a while. Finally we adiabatically reduce the rotation rate well below the critical frequency and detect the atoms in the rotating and non-rotating basis.  In the ideal case, all atoms should be found in one or the other. The probability of these two outcomes depends on the evolution after the non-adiabatic shift and should provide a measure of the angular momentum of the system. Importantly, the rate at which these probabilities change with the angular momentum should be $N$ times faster than if we used the usual Sagnac setup where the atoms are not entangled, but each is put into a superposition of going each way round the loop. This enhancement of the frequency of the interference fringes due to entanglement has already been observed in optical systems \cite{Mitchell2004a, Nagata2007a}. It offers the possibility of making measurements with enhanced sensitivity right up to the Heisenberg limit.

\section{Decoherence}

So far we have considered only NOON states. The Fisher information for these states given by (\ref{fisherpure})  is $F_Q = N^2$, which means that they can reach the Heisenberg limit, $\Delta\phi \geq 1/N$ when there is no decoherence. However they are known to be very fragile and so quickly lose their advantage when any loss is present. For more than a small amount of loss, the case of unentangled atoms each put into a superposition of the modes $\alpha$ and $\beta$, outperforms the NOON state.

\begin{figure}[t]
  \includegraphics[height=.26\textheight]{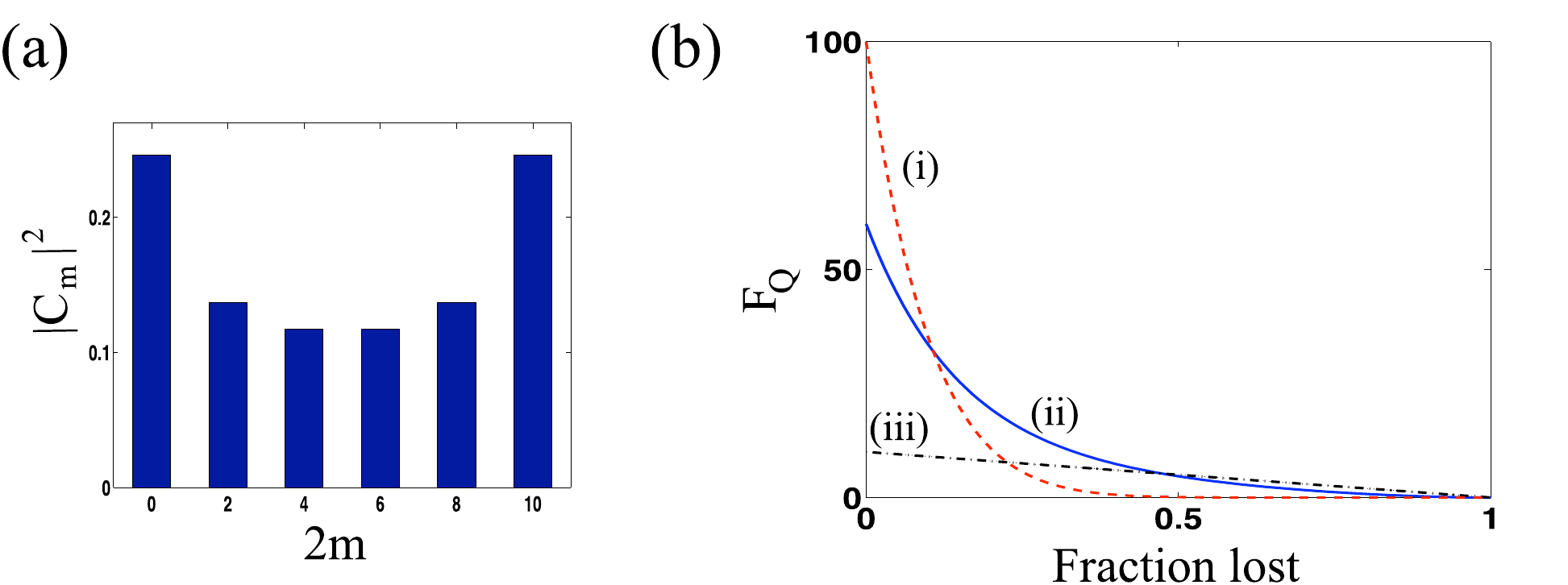}
  \caption{(a) Plot of the coefficients of a bat state with 10 atoms. (b) Comparison of the quantum Fisher information for three different states (each initially with 10 atoms) as a function of the fraction of particles lost: (i) NOON state, (ii) the bat state shown in (a), and (iii) the case where the atoms are not entangled with one another and each is in an equal superposition of modes $\alpha$ and $\beta$.} \label{fig:bat}
\end{figure}

However, it is possible to find states that embody the best aspects of each of these cases, i.e. states that outperform the precision of unentangled particles and yet don't suffer from the same fragility as NOON states. One such example, known as a bat state, is formed by passing a dual Fock state, $|\psi\rangle =|N\rangle|N\rangle$ through a 50:50 beam splitter \cite{Holland1993a, Dunningham2002a}. It has the form $|\psi\rangle = \sum_{m=0}^{N} C_m |2m\rangle |2N-2m\rangle$ where,
\begin{eqnarray}
C_m = \frac{\sqrt{(2m)! (2N-2m)!}}{2^N m!(N-m)!}.
\end{eqnarray}
These coefficients are plotted in Figure~\ref{fig:bat}(a). We see that the state is quite like a NOON state in that it has a large number variance. This means that it has a large Fisher information and hence can measure phase shifts with high precision. However, unlike a NOON state, if we detect which mode one of the atoms is in, this does not betray which mode all the remaining atoms are in. In other words, the bat state should be much more robust to the loss of atoms than the NOON state.

In Figure~\ref{fig:bat}(b) we have plotted the Fisher information for these three different states initially containing 10 atoms as a function of the fraction of particles lost. For simplicity, we have assumed that there is equal loss for both of the modes. We see that in the lossless case, the NOON state has the highest Fisher information, as expected. However, for moderate loss, the bat state outperforms both the NOON and unentangled states. These differences become even more dramatic as the number of particles is increased.

Since loss is inevitable in any realistic system, it would seem that it would be very useful if we could create bat states with rotating matter waves. A scheme to do this was recently proposed in \cite{Cooper2010a}.  However due the highly squeezed state required to create the bat state and the unwanted effects of particle interactions in the scheme, this matter wave gyroscope is limited to a small number of particles thereby limiting its practical precision capabilities.  So the question is, can we find practical ways for creating other entangled states that are well-suited to metrology in the presence of loss? This is still very much a new field of enquiry, however two intriguing examples show how this is not only possible, but that we can do even better than bat states. This is achieved by reducing the trapping potential down to one or two dimensions. In the following two sections we review these two cases.

\section{Two-dimensional pancake trap}

Dagnino {\it et al.} recently demonstrated a very interesting model for creating entangled atoms in a rotating potential \cite{Dagnino2009a}. They considered a two-dimensional `pancake' shaped potential. One of the dimensions can be removed in practice by taking the trapping frequency in (say) the $z$-direction to be much greater than trapping frequencies in the $x$ and $y$ directions, i.e. $\omega_z \gg \omega_{x,y}$. If $\hbar\omega_z$ is sufficiently large compared with the interaction energy, the dynamics along $z$ are frozen and the gas is effectively two-dimensional at sufficiently low temperatures.

The Hamiltonian for this system is,
\begin{eqnarray}
H = \sum_{i=1}^{N}\left(-\frac{\hbar^2}{2M}\nabla_i^2 +\frac{1}{2}M\omega_{xy}^2(x^2+y^2) + \frac{1}{2}M\omega_{z}^2 z_i^2 -\Omega L_{zi} +V_{i} \right)
+ \frac{g\hbar^2}{2M}\sum_{j\neq k}\delta(\vec{r}_j - \vec{r}_i), \label{Hpancake}
\end{eqnarray}
where the sums are over the particles, $M$ is the mass of an atom, $\Omega$ is the rate of rotation of the potential, $L_{zi}$ is the z-component of the angular momentum of particle $i$, and $\omega_{xy}$ is the radial frequency of the trap in the $xy$ plane. The interactions between atoms are parameterized by the unitless quantity, $g = a\sqrt{8\pi M\omega_z/\hbar}$, where $a$ is the 3D scattering length. The asymmetry is given by $V_{i} = 2AM\omega_{xy}(x^2 - y^2)$, where the coefficient $A$ measures the strength of the asymmetry.

When $A\ll1$ and $\Omega\approx\omega_{xy}$, the single particle energy levels are grouped into Landau levels separated by energy $\hbar(\Omega +\omega_{xy})$. This means that if the interaction energy is small compared with this spacing, to a good approximation, the system is restricted to the lowest Landau level \cite{Morris2006a}. A basis for the single particle states of the lowest Landau level for $A=0$ is,
\begin{eqnarray}
\psi_m = \sqrt{\frac{1}{m!\pi}\left(\frac{M\omega_{xy}}{\hbar^2}\right)^{m+1}} (x+iy)^m e^{-M\omega_{xy}(x^2+y^2)/2\hbar}, \label{LLLstate}
\end{eqnarray}
where $m\geq0$ is an integer. Using this basis, we can rewrite the Hamiltonian~(\ref{Hpancake}) in second quantised form,
\begin{eqnarray}
\hat{H} = \hbar\omega_{xy}\hat{N} &+& \hbar(\omega_{xy} - \Omega)\hat{L} +\frac{\hbar \omega_{xy}A}{2}\sum_m \left(\sqrt{(m+1)(m+2)}a_{m}^{\dag}a_{m+2} +\sqrt{m(m-2)}a_{m}^{\dag}a_{m-2}  \right) \nonumber \\
&+& \frac{\hbar \omega_{xy}g}{4\pi}\sum_{m_1,m_2} \sum_{n_1,n_2} \frac{\delta_{m_1+m_2,n_1+n_2}}{\sqrt{m_1!m_2!n_1!n_2!}}\frac{(m_1 +m_2)!}{2^{m_1+m_2}} a_{m_1}^{\dag}a_{m_2}^{\dag}a_{n_1}a_{n_2}, \label{secondquant}
\end{eqnarray}
where $a_j$ is the annihilation operator for a particle in the state $\psi_j$ given by (\ref{LLLstate}), and $\hat{N} = \sum_m a_{m}^{\dag}a_{m}$ and $\hat{L} = \sum_m m a_{m}^{\dag}a_{m}$ are respectively the number operator and the total $z$-component angular momentum operator.

The spectrum of energy levels for different parameters can be found by directly diagonalising (\ref{secondquant}). In the case that the potential is symmetric (i.e. $A=0$), the two lowest energy levels become degenerate at some critical rate of rotation, $\Omega_c = 1- gN/(8\pi)$. If we include some small asymmetry, this degeneracy is lifted and there is an anti-crossing at the critical frequency. This should be compared with the two other geometries we have discussed: the lattice loop and the continuous loop. In the former, the degeneracy was lifted by including interactions between the atoms and in the latter it was lifted by including a small barrier into the loop.

\begin{figure}[t]
  \includegraphics[height=.25\textheight]{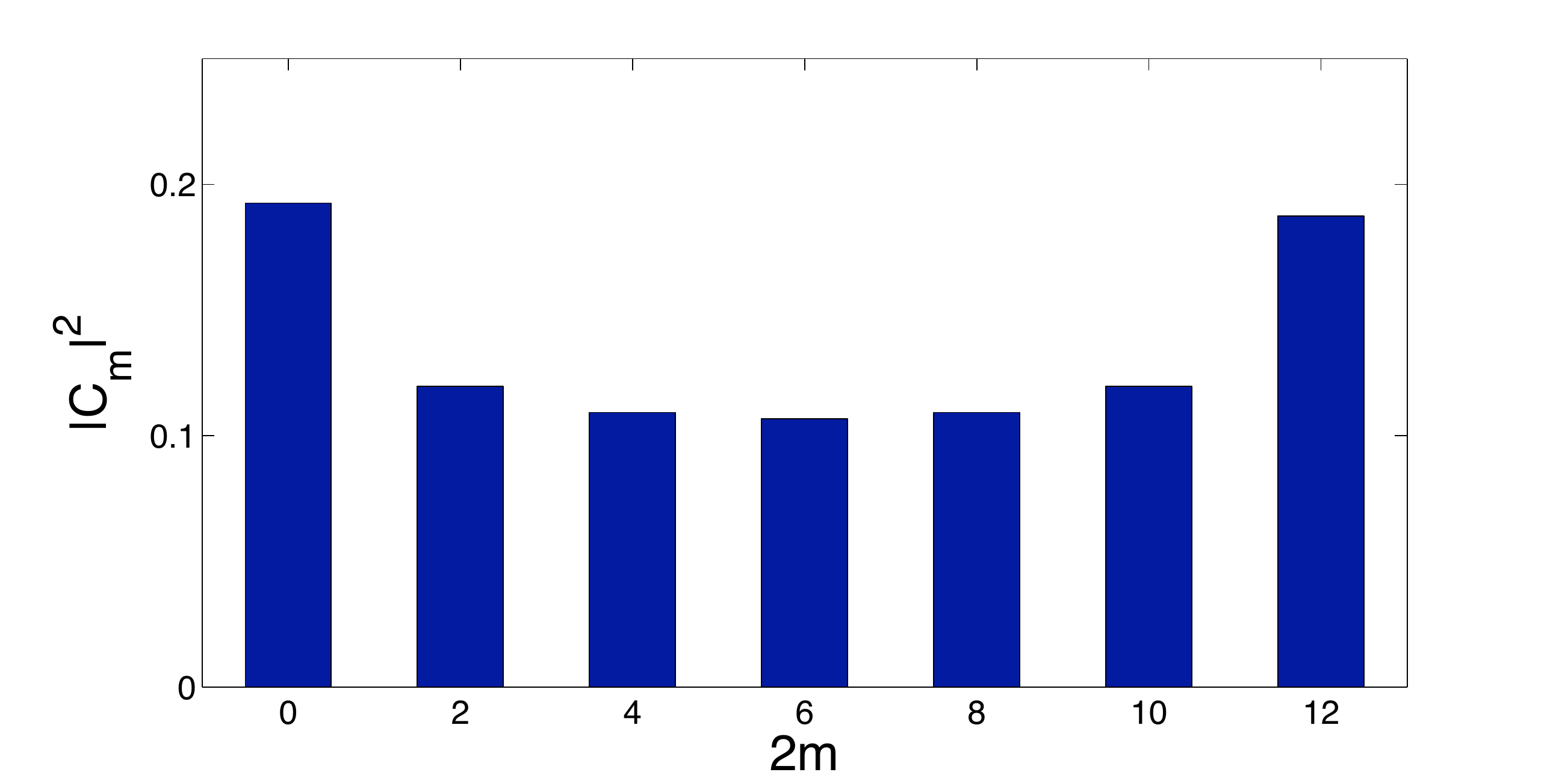}
  \caption{The ground state coefficients of the asymmetric 2D potential at the critical rotation rate, where the state has the form of Eq.~(\ref{2dstate}).
  The parameters chosen for this plot are  $N=12$, $g=0.5$ and $A=0.03$.} \label{fig:pancake}
\end{figure}

At the critical point, the ground state of the system (for even $N$) can be written approximately as 
\begin{eqnarray}
|\psi\rangle = \sum_{m=0}^{N/2} C_m |2m\rangle |N-2m\rangle, \label{2dstate}
\end{eqnarray}
 where the first ket represents an even parity state that is approximately a coherent superposition of the states $\psi_0$ and $\psi_2$, while the second ket is an odd parity state well-approximated by $\psi_1$ \cite{Dagnino2009a}. The coefficients of this ground state at the critical rotation rate are shown in Figure~\ref{fig:pancake} for the parameters $N=12$, $g=0.5$ and $A=0.03$. We see that the state is qualitatively similar to the bat state shown in Figure~\ref{fig:bat}(a). In particular, it has a large number variance and so has a large Fisher information meaning that it should be able to detect phase shifts very accurately. Also, the fact that the detection of a particle in one of the two particular modes that make up the state does not destroy the superposition means that it should be relatively robust to loss.

Overall, this seems to be a relatively straightforward scheme for generating an entangled state that is well-suited to metrology in the presence of loss. Developing a full scheme for how this system could be exploited as a quantum-limited gyroscope as well as analysing the behaviour of the state in the presence of decoherence are interesting directions for future work.

\section{One-dimensional loop and the Tonks-Girardeau limit}

Another promising example occurs in a one dimension ring trap \cite{Hallwood2010, CooperUnpub}. Suppose we have a ring potential with a moving barrier, similar to that shown in Figure~\ref{fig:continuous}(a), but where the potentials are so tight in the radial direction that we are left with only the angular, $\theta$, dimension. If the interactions between the atoms are strong, we enter the so-called Tonks-Girardeau (TG) regime where the atoms cannot pass one another.

The Hamiltonian that describes this system is
\begin{eqnarray}
H \! = \! \sum_{i=1}^N \left[ \frac{\hbar^2}{2M}\!\!\left( \!-i\frac{\partial}{\partial x_i} \!- \!\frac{\Omega}{L} \right)^2 \!\!\!+\!b \delta(x_i) \!+\! {g} \sum_{i<j}^N \delta(x_i\!\!-\!\!x_j) \right] ,
\label{eq:ham1}
\end{eqnarray}
where the sum is over all $N$ atoms of mass $M$. The circumference of the ring is $L$, which gives $x=\theta L/2\pi$ as the atom's position on the circumference of the loop and ${g}$ is the effective one-dimensional interaction strength between the atoms. The smallest nonzero kinetic energy of a single particle $E_0 = 2 \pi^2 \hbar^2/(M L^2)$ provides a natural unit of energy for this system.
The barrier is assumed to be narrow, so it can be described by a $\delta$-function and has strength $b$. The barrier stirs the atoms in the ring with tangential velocity  $v=\hbar\Omega/(ML)$ along the circumference of the ring and the Hamiltonian~(\ref{eq:ham1}) is formulated in the co-rotating frame of reference. 

It is convenient to write the Hamiltonian in the second quantized form using angular momentum operators,
\begin{eqnarray}
\hat{\Psi}(x) =\frac{1}{\sqrt{L}}\sum_k e^{i2\pi kx/L} \hat{a}_k,
\end{eqnarray} 
where $\hat{a}^{\dag}_k$ and $\hat{a}_k$ create and destroy an atom with angular momentum $k\hbar$, respectively. In this representation, the Hamiltonian is,
\begin{eqnarray}
\tilde{H}&=& \sum_{k} E_0 \left( k-\frac{\Omega}{2\pi} \right)^2 \hat{a}^{\dag}_k \hat{a}_k + \frac{b}{L} \sum_{k_1,k_2} \hat{a}^{\dag}_{k_1} \hat{a}_{k_2} + \frac{g}{2L} \sum_{k_1,k_2,q}  \hat{a}^{\dag}_{k_1} \hat{a}^{\dag}_{k_2}  \hat{a}_{k_1-q} \hat{a}_{k_2+q},
\label{eq:H3}
\end{eqnarray} 
where the sums are over all quantized angular momentum modes, $k\hbar$. Again the ground state and energy spectrum can be found by diagonalising this Hamiltonian, however, the Hilbert space is infinite so a truncated momentum basis has to be used. This means that the interaction strength needs to be rescaled as discussed in \cite{Hallwood2010} . 

Much can be understood about the system by considering the terms in the Hamiltonian. The kinetic energy term and the interaction term both conserve the total angular momentum and it is only the barrier term that couples states of different total angular momentum. When the barrier is rotated at a rate $\Omega=\pi$, the energy of the 0 and $\hbar$ angular momentum modes become degenerate, as described in section~\ref{sec:lattice_ring}. In the non-interacting case this means states with 0 to $N\hbar$ total angular momentum become degenerate. When interactions are introduced the degeneracy is lifted except the 0 and $N\hbar$ states and at a certain interaction strength, and with a finite barrier height, the ground state becomes similar to a NOON state. Increase the interactions further and the NOON state is lost, however the ground state is still a superposition of 0 to $N\hbar$ total angular momentum, because the interaction does not couple states of different total angular momentum. What we find is the states above this certain interaction still produce measurements with the same precision as the NOON state, however, their robustness is far better.

\begin{figure}[t]
  \includegraphics[height=.32\textheight]{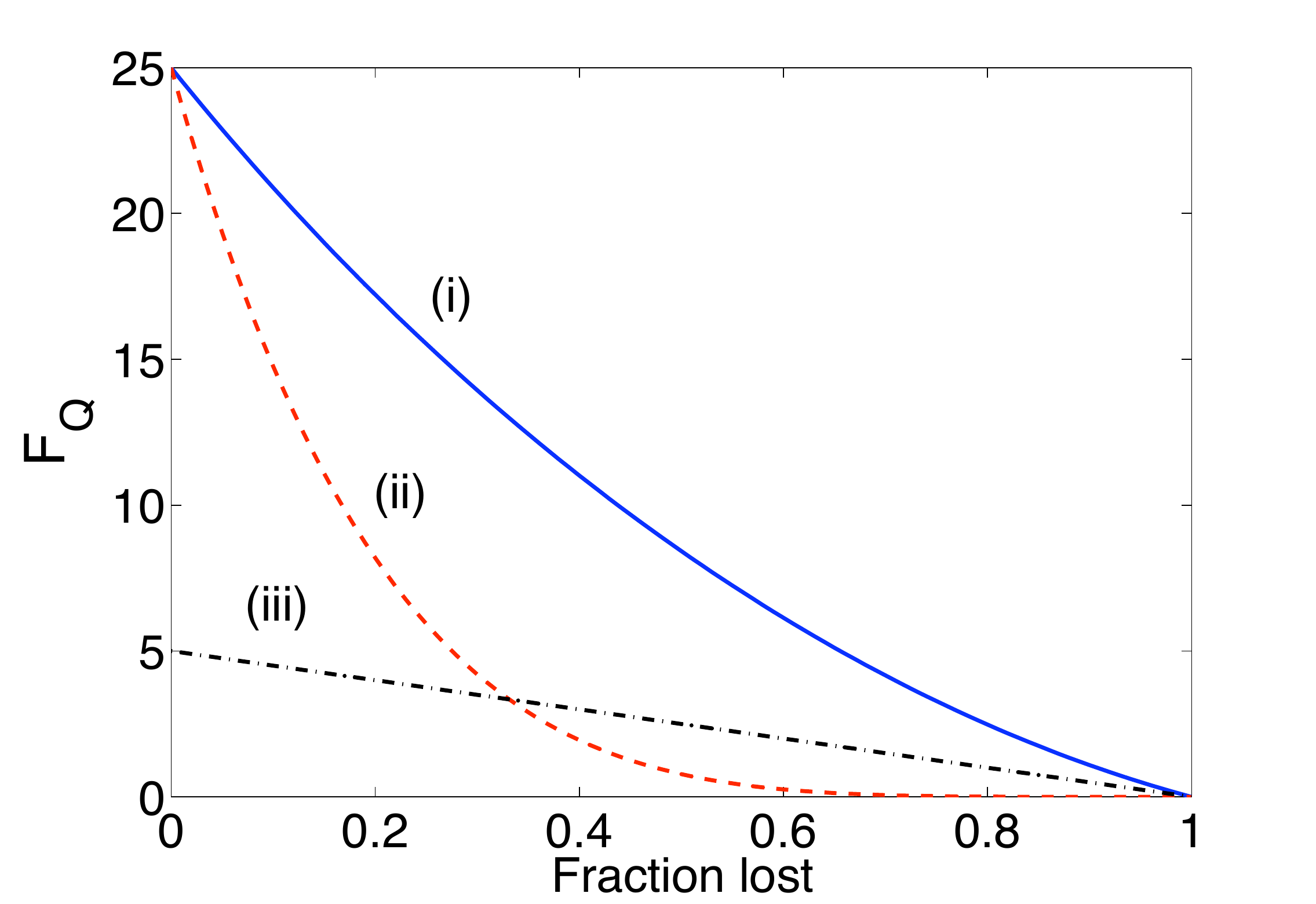}
  \caption{(i) The quantum Fisher information of a Tonks-Girardeau gas as a function of the fraction of atoms lost. This is compared with (ii) the NOON state and (iii) the unentangled state. In each case the state contains 5 atoms before loss. } \label{fig:TG}
\end{figure}

In the strongly interacting TG regime the state of the system looks very different from a NOON state. The single particle momentum distribution for the 0 and $N\hbar$ total angular momentum states are spread out and overlap each other. This means that if an atom is detected it cannot be discerned which of the total angular momentum states it came from and so it does not destroy the superposition. The performance of the TG superposition in the presence of loss is compared with the NOON and unentangled states in Figure~\ref{fig:TG} for $N=5$. The TG plot is for $b/L = 0.008E_0$ (i.e. a small barrier) and strong interactions, $g/L = 1085/(2\pi)E_0$, i.e. deep in the TG regime. The calculation was performed on a truncated basis with 18  angular momentum modes and the results do not change noticeably as the basis is increased. We see that the initial Fisher information is equal to that of the NOON state in the case without loss. When loss is considered, the reduction of the Fisher information is far slower and is always better than unentangled atoms. Another advantage of the system is the practical creation of the superposition. For the NOON state it was shown that even for the case of no decoherence it would be experimentally difficult to create for more than a few atoms due to the extreme precision required in the rotation rate~\cite{Hallwood2007a}. The TG superposition is far more stable to experimental imperfections and energy level spacings between the ground and excited states are far larger allowing more rapid creation of the superposition.

\section{Conclusions}
The idea of creating different entangled states by rotating matter waves in trapping potentials with different geometries is an exciting new area of research. There are already some interesting results for how entangled states that are suited to quantum metrology schemes could be generated relatively easily. There is still a lot to be done. It will be particularly interesting to see how different symmetry properties in the potential manifest themselves in different features of the entangled states. This may lead to quite general quantum state engineering proposals for rotating matter waves.
The models proposed here and their extensions may also prove to be valuable models for exploring the deeper connections between phase transitions, entanglement, and metrology.


\begin{theacknowledgments}
This work was supported by an RCUK Fellowship and the European Science Foundation through the EuroQUASAR programme EP/G028427/1.
\end{theacknowledgments}

\bibliographystyle{aipproc}   


\end{document}